\documentclass[osajnl,twocolumn,showpacs,superscriptaddress,10pt]{revtex4-1}
\usepackage{graphicx}
\usepackage{amssymb}
\usepackage{subfigure}
\usepackage{amsmath}
\usepackage{amsfonts}
\usepackage{bm}
\usepackage{color}
\newcommand{\vect}[1]{\ensuremath{\bm{\mathrm{#1}}}}

\graphicspath{{converted_graphics/}}
\begin{document}


\title{Optical integration of a real-valued function by measurement of a Stokes Parameter}

\author{Gabriela Barreto Lemos}
\email[]{gabriela.barreto.lemos@univie.ac.at}
\affiliation{Instituto de F\'{\i}sica, Universidade Federal do Rio de
Janeiro, Caixa Postal 68528, Rio de Janeiro, RJ 21941-972, Brazil}
\affiliation{Institute for Quantum Optics and Quantum Information, Boltzmanngasse 3, Vienna A-1090, Austria}
\affiliation{Vienna Center for Quantum Science and Technology (VCQ), Faculty of Physics, University of Vienna, A-1090 Vienna, Austria.}
\author{P. H. Souto Ribeiro}
\affiliation{Instituto de F\'{\i}sica, Universidade Federal do Rio de
Janeiro, Caixa Postal 68528, Rio de Janeiro, RJ 21941-972, Brazil}
\author{S. P. Walborn}
\affiliation{Instituto de F\'{\i}sica, Universidade Federal do Rio de
Janeiro, Caixa Postal 68528, Rio de Janeiro, RJ 21941-972, Brazil}

\begin{abstract}
We  experimentally implement an optical algorithm for integration of a real-valued bivariate function.  A user-defined function is encoded in the position-dependent phase of one of the polarization components of an optical beam.  The integral of this function is retrieved by measuring a Stokes parameter of the polarization.  We analyze the performance of the system
as an integration device.  
\end{abstract}



\maketitle
\section{Introduction}
Light is an extraordinary object for measurement, sensing, and communication applications
due to the extremely developed ability of manipulating its degrees of freedom, propagation speed, and very large 
bandwidth. Concerning information processing, there are a wide variety of methods 
\cite{goodman96,saleh91} based on incoherent light and coherent light, with applications to
signal processing, pattern recognition, and matrix algebra, among many others.

The potential of optical methods in real world computing has been under debate recently \cite{miller10,caulfield10,tucker10,caulfield10b,woods12}.  Key features of optical information processing is the possibility of parallelization using optical components such as lenses and holograms\cite{ozaktas96}, and the ability to perform the optical Fourier transform with diffraction or linear optical systems. Compared to digital electronic computing, optical computing presents a distinct paradigm for computation, which poses interesting fundamental questions of computational complexity \cite{woods09}, and provides alternative methods for sophisticated computational problems. In this regard, optical implementations have been presented for bounded NP-complete problems such as the traveling salesman \cite{oltean06,dolev07,shaked07,shaked07b,haist07a}, and artificial neural networks \cite{larger12,paquot12}.      
 \par
A simple optical method for integration of a non-negative valued function $f(x,y)$ is to encode the function into the intensity of an optical beam:  $I(x,y)=f(x,y)$.  The total intensity, given by the integral of $I(x,y)$, can then be retrieved directly by focusing the beam onto a detector  \cite{goodman96,saleh91}.  
\par
Here we wish to discuss a different method for optical integration of a real-valued function that is based on interference between two orthogonally polarized optical beams. By measuring the appropriate Stokes parameter, the integral 
\begin{equation}
J = \int\limits_{-x_\ell}^{x_\ell} \int\limits_{-y_\ell}^{y_\ell}  f(x,y) dx dy 
\label{eq:J}
\end{equation}
of a user-defined function $f(x,y)$ can be retrieved.  The function $f(x,y)$ may take on positive or negative values, however the integral must converge.  In our experiment, the function $f$ is programmed using a phase-only spatial light modulator.  An interesting feature of this optical method is that it does not depend upon the degree of spatial coherence of the beam, and requires only a very small coherence length. The readout of the integral $J$ depends entirely on the polarization interference, even though the function $f(x,y)$ is encoded into the spatial properties of the beam.


\section{Optical algorithm}  

\begin{figure}
  \begin{center}
\includegraphics[width=7cm]{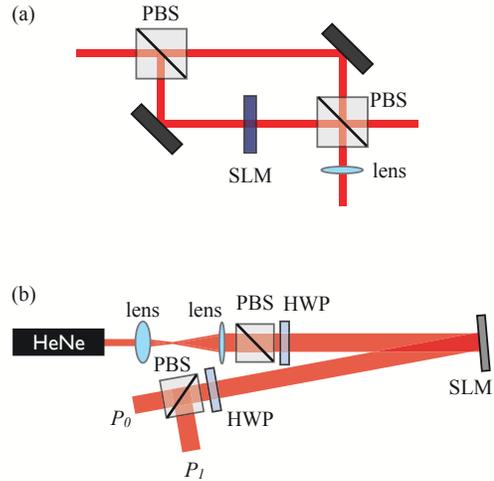}
  \caption{   (a) Schematic diagram of the optical integration algorithm. (b) Experimental setup for the integration of the function $f(x,y)$.}
\label{fig:setup}
\end{center}
  \end{figure}

The basic idea is illustrated in Fig. \ref{fig:setup} (a).  
Let us first consider the integration algorithm using coherent light.  
A laser with transverse profile $E(x,y)$ and polarized in the linear diagonal direction $\vect{e}_{+}$ is sent through a polarizing beam splitter (PBS), separating horizontal ($H$) and vertical ($V$) polarization components into separate arms of the interferometer. In the horizontal arm one imprints onto the field a phase $\exp[i a(x,y)]$ that is a function of the transverse coordinates $(x,y)$.  The two beams are then recombined at the output of a second PBS. Optical imaging systems are used to map the field at the phase element ($z=0$) onto the detection plane. The output of the interferometer is 

\begin{equation}
\vect{E}_{\mathrm{o}}(x,y)=\frac{E(x,y)}{\sqrt{2}}\left(e^{ia(x,y)}\vect{e}_{H} + \vect{e}_{V}\right),
\end{equation}
which can be rewritten as
\begin{equation}
\vect{E}_{\mathrm{o}}(x,y)=E(x,y)\left( \frac{1+e^{ia(x,y)}}{2}\vect{e}_{+} + \frac{1-e^{ia(x,y)}}{2}\vect{e}_{-} \right),
\end{equation}
where $\vect{e}_{\pm}$ are the linear diagonal ($+$) and antidiagonal ($-$) polarization unit vectors.  {By focusing the incoming field onto a finite-size detector, one can 
realize the area integrating detection of the $\pm$ polarization field components, obtaining the intensities }
\begin{equation}\label{eq:intensity}
I_\pm=\iint |E(x,y)|^2  \left \{ \frac{1}{2} \pm \frac{1}{4}  \left ( e^{ia(x,y)}+ e^{-ia(x,y)}\right )\right \} dx \; dy,
\end{equation}
and
\begin{equation}\label{eq:intcos}
I_\pm=\iint |E(x,y)|^2  \frac{1}{2} \left \{ 1 \pm \cos\left [a(x,y) \right ] \right \}dx\; dy  .
\end{equation}
{ The difference $D=I_+-I_-$ in intensity measurements gives the Stokes parameter $S_2$} \cite{hecht}:  
If $a(x,y) = \arccos [f(x,y)]$, 
\begin{equation}\label{eq:dif1}
D =\iint |E(x,y)|^2 f(x,y) dx \; dy.
\end{equation}
{Assuming that $f(x,y)$ is zero outside $-x_\ell < x < x_\ell$ and $-y_\ell < y < y_\ell$, and that $|E(x,y)|^2 \approx |E(x_0,y_0)|^2$ is constant inside this region, the intensity difference, $D$, is  proportional to the integral, $J$, defined in Eq. \eqref{eq:J}}:
\begin{equation}
D \approx |E(x_0,y_0)|^2 J.
\label{eq:integration}
\end{equation}
{Hence, $J \propto D/T$, where $T=I_++I_-$ is the total detected intensity of light. 
This provides a simple optical method to calculate the integral of limited real two-dimensional functions}. 
\par
In order to estimate the uncertainty $\delta_J $ in the value of the integral $J$, let us consider only intensity fluctuation errors, and that the intensity of each polarization component is given by $I_{\pm}=g \langle n_{\pm} \rangle$, where  $\langle n_{\pm} \rangle$ is the mean number of photons, and $g$ is a constant. Assuming Poissonian statistics at the shot noise level, with $\delta_{\langle n \rangle} = \langle n \rangle^{1/2}$, we have $\delta_D =  \delta_T = (g T)^{1/2}$, which gives
 \begin{equation}
 \delta_J \propto  (g/T)^{1/2}(1+D^/T) \leq 2(g/T)^{1/2}.
 \label{eq:error}
 \end{equation}  
Thus, the uncertainty in $J$ scales as the square root of the inverse of the intensity of light. 
\section{Experiment}

To implement a proof of principle realization of the integration algorithm we polarize a $633$nm He-Ne laser beam in the diagonal direction, and direct it onto a Holoeye Spatial Light Modulador (SLM), as illustrated in Fig. 
\ref{fig:setup} b). Reflecting the beam upon the high-definition LCD screen of this device imprints a programmable phase $\exp[ia(x,y)]$ on the horizontal polarization component of the beam, while the phase of the vertical polarization component remains unchanged. This polarization dependent action of the SLM circumvents the need to build an interferometer using a PBS  (as shown schematically in Fig.\ref{fig:setup}(a)), resulting in a very stable setup. Moreover, the coherence length required for coherent operation is very small,
as decoherence effects induced by the SLM are practically negligible\cite{Lemos_tomoproqSLM}. Two spherical lenses are used to map the reflected light field onto a power meter. By placing a half-wave plate (HWP) and a PBS  in front of the area integrating detector one can measure $I_+$ and $I_-$, 
Eq.\eqref{eq:intensity}, corresponding to each of the two linear diagonal polarization states, 
$(\vect{e}_{\pm}=\vect{e}_{H}\pm \vect{e}_{V})/\sqrt{2}$. 
  
Before the SLM, the beam is expanded and collimated so that the  field amplitude is approximately constant in the region over which the phase function is imprinted. By programming the SLM to imprint the phase function $a(x,y) = \arccos[f(x,y)]$ onto the horizontal component of the beam, the difference between the intensity measurements, $I_+$ and $I_-$, yields the result of the integration Eq.\eqref{eq:integration}. The balance between $H$ and $V$ polarizations in the SLM is made with
half waveplate HWP1, and the projection onto diagonal basis is adjusted using half waveplate HWP2.  

\section{Calibration}

From  Eqs. (\ref{eq:intcos}) and (\ref{eq:dif1}) we notice that when
$f(x,y) = 0$ and consequently $J=0$, the phase imprinted by the SLM should be
$a(x,y)=\pi/2$. Complete cancellation of the constant terms in (5) occurs only when the background counts for $I_\pm$ are exactly equal.  In practice, this is never true, due to intensity fluctuations of the laser beam and imperfections in the optics.  However, the systematic errors result in a constant background, and this can be compensated through a
calibration procedure.
Another parameter that must be calibrated is the value $|E(x_0,y_0)|^2$ in Eq. (6).  Since the intensity of the beam is approximately constant in the region where $f(x,y)$ is nonzero, this parameter is proportional to the total intensity.

We can take both of these calibration issues into account by assuming that the difference in intensity measurements is given by 
\begin{equation}\label{dif}
D=I_+ - I_- =  B + A J,
\end{equation}
where $J$ is the value of the integral \eqref{eq:J}, and $A$ and $B$ are constants that depend upon the experimental setup, as discussed above. Ideally, $B=0$ and 
$A = T$. We can experimentally 
determine the values of $A$ and $B$ performing a 
calibration procedure, as explained above.

To calibrate the device and determine $A$ and $B$, we use a set of functions for which we already know the value 
of the integral. For example, for one-dimensional integrals our test functions will be 
the Gaussian functions
\begin{equation}
f_{\rm G}(x,y) =  e^{-(x/\sigma)^2} R(y, y_\ell),
\label{eq:g}
\end{equation} 
where $R(y,y_\ell)$ is a unity-height step function of width $y_\ell$ (pixels) around the origin. 
The analytical integration of $f_{\rm G}(x,y)$ gives 
$J(\sigma,y_\ell=200 \;{\rm})= 400\sqrt{\pi}\sigma$.
We measure $D[J(\sigma,y_\ell=200)]$ for several values of the width $\sigma$, 
and we obtain $A$ and $B$ from relation $D[J(\sigma,200)] = B + A 400\sqrt{\pi}\sigma$,
given by Eq. \eqref{dif}. 
The measurement results are shown in the inset of Fig.\ref{fig:hgx}.  By varying $\sigma$ and fitting the experimental data, one obtains the calibration parameters $A$ and $B$.  In this case, we obtained $A=(2.8 \pm0.2)\times10^{-3}\mu W/{\rm pixel}^2$ and $B=(-69 \pm 3) \mu W$. We can then use these calibration parameters to associate the desired integration result with $D$, \textit{i.e.} $J=(D-B)/A$. 
\par
The same setup can be used to integrate bivariate functions which are not separable in the two spatial dimensions. In this case, the calibration of the parameters $A$ and $B$ can be determined by calculating two-dimensional integrals of a test function such as 
\begin{equation}
f_{\rm G2}(x,y) = \exp\left(-\frac{x^2+y^2}{\sigma^2}\right) R(x, x_\ell, y, y_\ell).
\end{equation}
Assuming that $\sigma$ is small enough compared to $x_\ell$ and $y_\ell$, so that the integration limits can be extended to infinity, $\int_{-\infty}^\infty f_{\rm G2}(x,y) dx\; dy$ gives $J(\sigma)=\pi \sigma^2$. 
 The inset of Fig.\ref{gauss-hg4} shows a plot of $D$ as a function of $\sigma$, from which we obtain the values $A=(30.3 \pm 0.6)\times 10^{-4} \mu W/{\rm pixel ^2}$ and $B=(25 \pm 1) \mu W$.  
\par

\section{Integration of 1D and 2D functions}

\begin{figure}
  \begin{center}
\includegraphics[width=7cm]{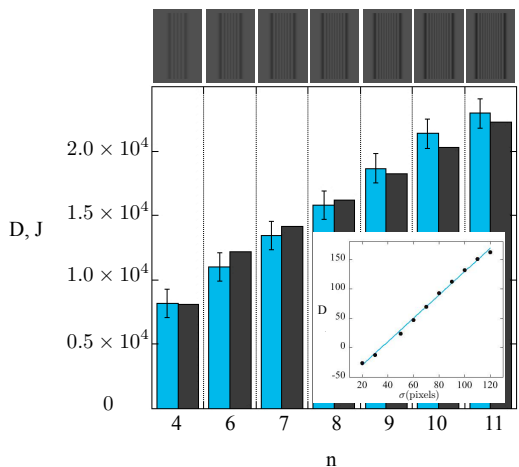}
  \caption{Optical integration(blue) and analytical result (black) for the function $h(x,y)$ given in Eq. \eqref{hg}. The inset shows the linear fit used to obtain the calibration parameters $A$ and $B$ for test function $f_{\rm G}(x,y)$ in Eq. \eqref{eq:g}.   The figures along the top show the greyscale image of the phase used on the SLM in each case.}
\label{fig:hgx}
\end{center}
  \end{figure}

After calibrating the setup, we tested the integration algorithm, estimating the integral of the function 
\begin{equation}
h_{n}(x,y) = \frac{ [H_n(x/20)]^2 e^{-(x/20)^2}}{2^n7 (n - 1)!}R(y,200),
\label{hg}
\end{equation}
where $ H_n(x)$ is the Hermite polynomial of order $n$. The integral of this function can be calculated analytically:
\begin{equation}
\int\limits_{-x_\ell}^{x_\ell} \int\limits_{-y_\ell}^{y_\ell}  H_{n}(x,y) dx \;dy \approx \frac{20}{7} 400 \sqrt{\pi} n, 
\label{eq:hgres}
\end{equation}
where $x_\ell$ is large enough so that the integration limits can be set to infinity, and the error due to the approximation is much smaller than the experimental errors.
In Fig. \ref{fig:hgx} we compare the result of the optical integration (blue bars) with the analytical result \eqref{eq:hgres} (black bars), for several orders $n$ of the Hermite polynomial. In the determination of the error bars, we considered the uncertainty in the measurement of $D$, corresponding to the fluctuations in 
the intensity of the He-Ne laser.  A single detector was used to measure each intensity $I_{\pm}$ separately, by
switching the polarization with a half waveplate. Thus, we expect a small fluctuation of the values of $A$ and $B$ due to changes of the laser intensity over time, even with the calibration technique described above. We believe this to be our primary source of experimental error. Proper balanced measurement of $I_{\pm}$ with two detectors can in principle  reduce this uncertainty to the shot noise level. 
\par
In order to analyze the method for 2D functions, we estimated the integral
\begin{equation}
h^\prime_{n}(x,y)=\frac{1}{30^n (n-1)!} H_n\left(\frac{xy}{20}\right)\exp\left(-\frac{x^2}{20}-\frac{y^2}{30}\right),
\label{eq:hprime}
\end{equation}
and 
\begin{equation}
S_\nu(x,y)=\sin\left(\frac{ \nu xy}{1000}\right)\exp\left(-\frac{x^2+y^2}{100}\right),
\label{eq:S}
\end{equation}
using our method and comparing the result with the analytical solution.

\begin{figure}
  \begin{center}
\includegraphics[width=7cm]{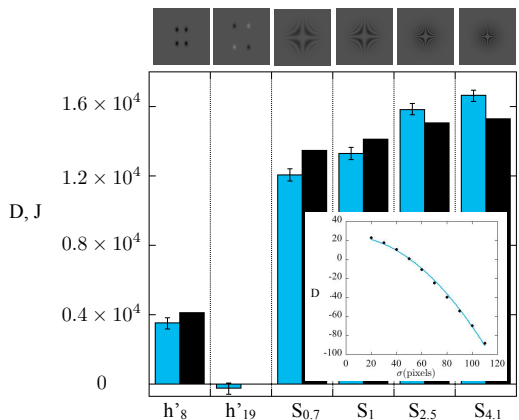}
  \caption{   
Optical integration(blue) and value for the analytical integration (black) of different two-dimensional functions $h^\prime(x,y)$ and $S(x,y)$ given in Eqs. \eqref{eq:hprime} and \eqref{eq:S}. The inset shows the quadratic fit used to obtain the calibration parameters $A$ and $B$. The figures along the top show the greyscale image of the phase used on the SLM in each case.}
\label{gauss-hg4}
\end{center}
  \end{figure}

Fig. \ref{gauss-hg4} shows the experimental results for several values of the order $n$ of the Hermite polynomial in \eqref{eq:hprime} and frequency $\nu$ in \eqref{eq:S}.  
The experimental results are represented by the blue bars and theoretical prediction by black bars.

In both 1D and 2D cases, the agreement is reasonable, serving as a proof-of-principle demonstration of the optical integration method. There are three main sources of noise
and systematic errors in this system. One source is related to the SLM characteristics
(noise-phase fluctuations) and finite resolution(systematic). A second type comes from imperfect polarization optics, and
a third type of noise comes from the intensity fluctuations of the light field. 

The SLM introduces a phase noise, which reduces the visibility of the polarization interference.  This type of effect has been analyzed in \cite{Lemos_tomoproqSLM}.  The main effect of this noise in terms of the integration method is to decrease the signal to noise ratio, so that the overall intensity must be increased to achieve a certain precision in the evaluation of the integral. In terms of the pixelization of the SLM, we can say that for functions with
spatial oscillations having a wavelength much larger that the SLM pixel size, the noise
introduced by the finite resolution of the SLM is very small.

Finally, the noise coming from the intensity fluctuations of the light beam
is dominant in our set-up. The beam is not at the shot-noise level, and we do 
not perform a balanced detection. We hope we will improve the performance from
this point of view in a future realization, to demonstrate optical integration 
with much smaller errors.

\section{Spatial coherence}

We have shown that the optical method can be used to integrate a real-valued function encoded in a spatial dependent phase of an optical beam. However, the method is not dependent on the transverse spatial coherence
of the light beam used. One would expect that transverse spatial coherence should be required,
because the method relies on the interference between the two polarization field modes. However,
because of the SLM property of modulating only one polarization mode, the transverse modes are never
separated and propagate together. Moreover, the image of the SLM plane is projected onto the
detection plane, so that the spatial phase modulation does not affect the propagation from the
SLM to the detection plane. Finally, the detector integrates over the whole transverse profile.
In a few words, the interference effect observed consists of interference point by point within the
wavefront. What is actually necessary is polarization coherence, between the $H$ and $V$ components of the light field.
        
\section{Conclusion}

We have presented and experimentally demonstrated an optical method for integration of a real-valued bivariate function.  Though the function is encoded into the spatial profile of an optical beam, the value of the integral is retrieved by measuring the Stokes parameter of the polarization of the beam.  In our proof-of-principle experiment, several functions were programmed into the position-dependent phase of an optical beam using a spatial light modulator.  The precision of the optical algorithm increases with the square root of the intensity of the beam. The method does not depend upon the spatial coherence of the beam, but rather relies on the degree of polarization. We expect our results to stimulate other applications in optical information processing.    

 \begin{acknowledgments}
 Financial support was provided by Brazilian agencies CNPq, 
CAPES, FAPERJ, and the INCT-National Institutes for Science and Technology - Quantum Information.
 \end{acknowledgments}
\bibliographystyle{osajnl}

\end{document}